\documentclass[usenatbib]{mn2e}                                                 
\usepackage{graphicx}                                                           
\usepackage{aas_macros}                                                         
\usepackage{amsfonts}                                                           
\usepackage{bm}  %allows bold fonts for Greek letters                           
\usepackage{times}                                                              
\begin{document}                                                                
                          
%%%%%%%%%%%%%%%%%%%%%%%%%%%%%%%%%%%%%%%%%%%%%%%%%%%%%%%%%%%%%%%%%%%%%%%%       
%%% Extra definitions                                                           
                                                                                
\def\beqn{\begin{equation}}
\def\eeqn{\begin{equation}}
\def\beqnarray{\begin{eqnarray}}                                                
\def\eeqnarray{\end{eqnarray}}                                                  
\def\nn{\nonumber}
\def\norml{{\bf{\hat{l}}}}                                                      
\def\norms{{\bf{\hat{s}}}}                                                      
\def\vecv{{\bf{v}}}                                                             
\def\deg{\hbox{$\null^\circ$}}                                                  
                                  
\def\Msun{\,\textrm {M}_\odot}                                              
\def\kpc{\,{\rm kpc}}                                                           
\def\Myr{\,{\rm Myr}}                                                           
\def\kms{\,{\rm km}\,{\rm s}^{-1}}
                   
\def\ud{\mathrm{d}}                                                             
%%%%%%%%%%%%%%%%%%%%%%%%%%%%%%%%%%%%%%%%%%%%%%%%%%%%%%%%%%%%%%%%%%%%%%%%%       

\title[Inferring the dynamics of stellar streams via distance gradients]
{Inferring the dynamics of stellar streams via distance gradients}

\author[S.~Jin \& N.~Martin]
{Shoko Jin$^{1}$\thanks{e-mail: shoko@ari.uni-heidelberg.de} and Nicolas F. Martin$^{2}$\\
$^{1}$Astronomisches Rechen-Institut, Zentrum f\"ur Astronomie der Universit\"at Heidelberg, M\"onchhofstr. 12--14, D-69120, Heidelberg, Germany\\
$^{2}$Max-Planck-Institut f\"ur Astronomie, K\"onigstuhl 17, D-69117, Heidelberg, Germany}

\date{}

\maketitle

\begin{abstract}
We present a simple result in which the distance gradient along a
stream can be used to derive the transverse velocity (i.e. proper
motion) along it, if the line-of-sight velocity is also known.  We
show its application to a mock orbit to illustrate its validity and
usage.  For less extended objects, such as globular clusters and
satellite galaxies being tidally disrupted, the same result can be
applied in its small-angle approximation.  The procedure does not rely
on energy or angular momentum conservation and hence does not require
a Galactic model in order to deduce the local velocity vector of the
stream.

\end{abstract}


\section{Introduction}

In the last few years, our view of the Milky Way's stellar halo has
been altered by the pioneering work of the Sloan Digital Sky Survey
(SDSS) and continued analyses of its data; it is now abundantly clear
that the outskirts of the Galaxy provide the prime locations in which
to search for signs of ongoing accretion events onto our Galaxy, as
well as the remnants of such events, more progressed toward their
final dissolution into the Milky Way \citep{2006ApJ...642L.137B}.
Current survey data that enable such studies are mainly photometric,
while spectroscopic follow-up is often necessary to check, via its
kinematics, whether a photometrically selected stellar grouping is
truly an association.  Many stellar streams originating from either
globular clusters or dwarf galaxies have been identified in the
Galactic halo through such investigations.  The former can be
exemplified by the globular cluster Palomar~5 \citep{odenk01,odenk03},
while the most prominent example of the latter is the Sagittarius
stream \citep{1994Natur.370..194I}.  

Much of the recent attempts to understand the dynamics of these
objects, and thus also their ultimate fate, have come from their
modelling via numerical simulations that can now foster millions of
particles in a study of a single stellar system.  These works
illustrate how the tidal distortion and subsequently resulting
disruption lead to orbits of constituent stars that deviate from that
of the bound parental structure
\citep[e.g.][]{1997AJ....113..634I,dehnen04,2005ApJ...619..800J,2007MNRAS.381..987C,2007ApJ...659.1212M}.
Fully live numerical studies also allow for the possibility of
studying disrupting systems under the auspices of dynamical friction.
Simulations are therefore key to furthering our understanding of these
dynamically evolving stellar substructures, with improvements and
advancement continually being made on the incorporation of necessary
physics in the correct ratios to model better the systems under study.

It is, however, also interesting to reflect on the extent to which
simple geometric or analytic arguments can aid in these quests.
Previously, we showed that the radial velocity gradient along a stream
can be used to derive the transverse velocity (i.e. proper motion) as
a function of distance (\citealt{2007MNRAS.378L..64J}; see also
\citealt{2008MNRAS.386L..47B}), which in turn allows for the
determination of orbits for these streams.  The success of this method
relies on the radial velocity gradient being well determined, and
Binney showed its successful application to Palomar~5, by recovering
its observationally known properties.  The procedure presented in this
letter is a related and complementary one, applicable when the
distance information for stars in a stream is more abundant (or better
constrained) than that for radial velocities.  In the following
sections, we first present a simple result that can be used to
calculate the transverse velocity for a given point along a stellar
stream, using knowledge of the distance gradient and the radial
velocity there.  An application to a mock orbit is then used to
illustrate its usage.  We also provide a simple geometric
visualisation of how the same result can be reached for less extended
objects, such as dwarf galaxies that are being tidally disrupted.

\section{Constraining transverse motion through distance gradients}

\subsection{The method}
\label{subsec:general}

Let $d\norml$ be the heliocentric position vector to a
particular point on the stream, with unit line-of-sight vector
$\norml$ and heliocentric distance $d$, so that its line-of-sight
velocity, relative to the Galactic Standard of Rest, is given by $v_l
= \vecv.\norml$.  Differentiating the position vector with respect to
time, $t$, and using the chain rule results in the following:
\beqnarray
\label{eqn:gradd}
\vecv &=& \frac{\ud d}{\ud t}\norml + d \frac{\ud\norml}{\ud t}\nn\\
&=& \frac{\ud d}{\ud\chi}\frac{\ud\chi}{\ud t}\norml 
+ d\frac{\ud\norml}{\ud\chi}\frac{\ud\chi}{\ud t}\nn\\
&=& \frac{v_s}{d}\frac{\ud d}{\ud\chi}\norml + {v_s}{\norms}\,,
\eeqnarray
where $\chi$ is the angle along the stream, measured from some
fiducial point, and $\norms$ is the unit vector along the apparent
direction of motion of the stream at $\norml$, such that $\norms =
\ud\norml/\ud\chi$ and $\ud\chi/\ud t = \vecv.\norms/d = v_s/d$ from
the usual equation for the transverse velocity.  One possible method
for calculating $\norms$ is provided by \citet{2008MNRAS.383.1686J}.
By noting that the full velocity vector at any point along a stream is
given by $\vecv = v_l\norml + v_s\norms$, we obtain the following
relation:
\begin{equation}
\label{eqn:vs}
v_s = d v_l \left(\frac{\ud d}{\ud\chi}\right)^{-1}\,.
\end{equation}
This result can also be reached by taking the derivative of the
position vector with respect to $\chi$, rather than $t$.  We note that
equation~(5) of \cite{2008MNRAS.386L..47B} gives the same result, by
identifying the time derivative of the angle as $v_s/d$.

While equation~(\ref{eqn:vs}) does not require the value of the
Galactic potential at the location of interest as an input, it assumes
that the stars of the stream, whose distances are used as an
ingredient of the method, are part of the same moving group and that
the stellar trajectories follow the observed stream.  Under this
assumption, the method provides a means of calculating the stream's
local transverse velocity, thereby allowing the determination of all
components of the velocity vector.  In an ideal case, one would have
both distance and kinematic data on the stars.  However, if both
distances and line-of-sight velocities are known, then we could just
as easily calculate the transverse velocity through a method that
relies on the gradient in $v_l$ \citep{2007MNRAS.378L..64J}.  The
result presented in this letter is therefore most useful when the
distance gradient along a stream is well determined, and for
relatively small uncertainties in the stellar distances (e.g. for
RRLyrae or BHB stars), but when one's knowledge of the kinematics is
more limited.  On the other hand, if no radial velocity information is
available for the stars, one can still determine the ratio of the two
velocity components.  Hence the family of orbits to be derived for the
stream need only be a function of $v_l$.

Given the simple form of equation~(\ref{eqn:vs}), the error in the
derived transverse velocity is calculated very straight-forwardly
through the quadrature sum of the fractional errors in the other
quantities:
\begin{equation}
\label{eqn:err}
\frac{\Delta v_s}{v_s} = \left(\left(\frac{\Delta Q}{Q}\right)^2 + 
\left(\frac{\Delta d}{d}\right)^2 + 
\left(\frac{\Delta v_l}{v_l}\right)^2 \right)^{1/2}\;,
\end{equation}
where $Q = \ud d/\ud\chi$ and $\Delta x$ gives the magnitude of the
error, or uncertainty, in quantity $x$.  This warns us that if any of
the quantities are very small, then its fractional error will
significantly boost the error for $v_s$ and that, in particular, a
small distance gradient will cause the most damage, given its place in
the denominator of equation~(\ref{eqn:vs}).

\subsection{Example application}
\label{subsec:example}

\begin{figure}
\begin{center}
\includegraphics[width=0.3\textwidth, angle=270]{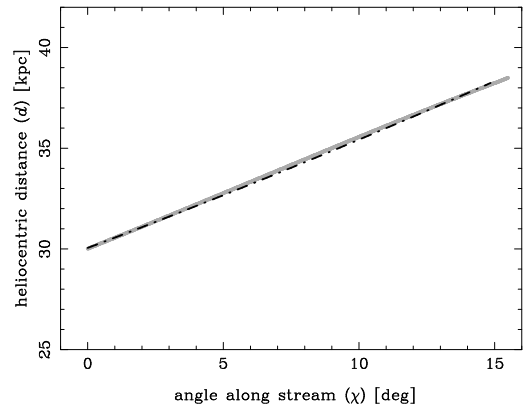} 
\caption{Distance as a function of angle along a small segment of a
  mock stream (solid grey line), overplotted with a fit deduced using
  a second order polynomial, $d = a + b\chi + c\chi^2$ (dot-dashed
  black line); the fit shown has $a = 30.1, b = 0.509, c =
  2.86\times10^{-3}$. $\chi=8.0\deg$ denotes the location where we
  choose to calculate the distance gradient in order to deduce the
  transverse velocity, and hence the location for which the initial
  conditions for the orbit calculation are subsequently
  defined.\label{fig:dchi}}
\end{center}
\end{figure}

\begin{figure}
\begin{center}
\includegraphics[width=0.4\textwidth, angle=0]{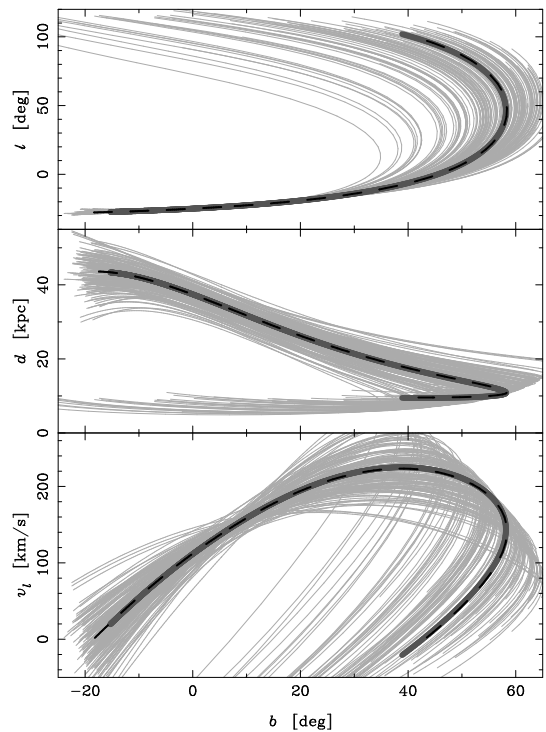}
\caption{Testing the recovery of the orbital parameters: Galactic
  longitude (top panel), heliocentric distance (middle panel) and
  line-of-sight velocities relative to the Galactic Standard of Rest
  (bottom panel) are shown as functions of Galactic latitude.
  Original data from the mock orbit are shown in dark grey (solid),
  whilst those for the recovered orbit are plotted in black (dashed).
  There are no significant differences between the two orbits over
  $70\deg$.  Also shown in light grey are 147 orbits for which artificial
  uncertainties were added to the orbital starting point for the
  distance, line-of-sight velocity and distance gradient; values for
  each parameter are drawn from a Gaussian distribution centred on
  the value characterising the correctly recovered orbit, with
  its $1\sigma$ deviation set to $7\%$ that of the distribution mean.
\label{fig:orbit_test}}
\end{center}
\end{figure}

Figure \ref{fig:dchi} shows the distance evolution along a mock orbit
(in grey), generated using the Milky Way model adopted by
\citet{1990ApJ...348..485P}.  The exercise here is to use the distance
gradient and a single radial velocity to deduce the transverse
velocity at a given point on the stream, and then attempt to recover
the input orbit.  Within this orbit, we place ourselves at a point
mid-way along the segment shown in Figure~\ref{fig:dchi}, at
$\chi=8.0\deg$.  The distance gradient is calculated by simply fitting
a second-order polynomial (also shown in the same figure by the
dot-dashed black line) and taking its derivative at the location of
interest\footnote{The choice of location is reasonably random, apart
  from avoiding regions with little distance variation; this would
  lead to unreliable values for $v_s$ as explained in the main text.}.
The heliocentric distance and line-of-sight velocity are $34.5\kpc$
and $137.9\kms$, respectively, and through the application of
equation~(\ref{eqn:vs}), we determine the transverse velocity to be
$149.4\kms$.  The directional vector $\norms$ is found to be
$(-0.014,-0.254,-0.967)$ in Cartesian Galactic coordinates, using
equation~(16) of \citet{2008MNRAS.383.1686J} on two locations,
separated from the central point by $5\deg$.  These allow us to
calculate the velocity vector, from which we then have the necessary
6-dimensional phase-space information to derive the orbit.  The
original orbit is properly recovered in this example, with the
Cartesian velocity vector being $(123.5, -93.4,-131.8)\kms$ in the
deduced orbit at the position specified above, where it was
$(123.7,-93.0,-130.8)\kms$ in the original orbit.  The small
deviations in the velocity vector components do not lead to any
noticeable difference in any of the orbital parameters over $70\deg$,
as shown by the similarity between the solid grey and dashed black
lines in Figure~\ref{fig:orbit_test}. Note that $\chi=8.0\deg$ in
Figure~\ref{fig:dchi} corresponds to a Galactic longitude and latitude
of $(\ell,b) = (-23.8,5.3)\deg$ in Figure~\ref{fig:orbit_test}.

The panels of Figure~\ref{fig:orbit_test} also show the results of a
simple exercise, whereby the parameters $d$, $v_l$ and $\ud d/\ud\chi$
at the orbital starting point were allowed to vary, in order to mimic
and ascertain the possible effects of observational uncertainties for
a real stellar stream.  Each of the three parameters were drawn from a
Gaussian distribution with the mean given by the initial conditions of
the correctly recovered orbit, and whose $1\sigma$ deviations were set
to $7\%$ of the value of the distribution mean.  In practice,
observational uncertainties are likely to be fractional for distances
($\sim7\%$ for RRLyrae and BHB stars) and absolute for velocities
($\sim10\kms$).  For the parameters of our mock orbit, $7\%$ of the
mean value corresponds to a realistic uncertainty in each of the
variables.  The resulting set of 147 light grey lines in each of the
panels therefore indicate the degree to which the recovered orbit
might differ from the original in cases where we might expect the
uncertainties for $d$, $v_l$ and $\ud d/\ud\chi$ to each be of the
order of a few percent.  As expected from this form of artificial
`degradation', the agreement of the sky positions is remarkably good
within $\pm20\deg$ of the orbital starting point, while the distance
tends to suffer a constant offset and the recovery of the
line-of-sight velocity is not as good.  Unsurprisingly, deviations
from the original orbit in all orbital parameters become larger with
increasing separation from the orbital starting point, with deviations
also manifesting themselves strongly near the turning point of the
orbit.

It is also necessary to add the usual cautionary note that streams do
not follow exact orbits.  However, the notion of quantifying their
transverse velocity in this way provides a useful tool with which the
motions of streams can be studied with limited kinematic information.
Figure~\ref{fig:orbit_test} also highlights that, in general, one
should preferentially apply the technique --- observations allowing
--- every $\sim20\deg$ along the stream, so as to follow better the
stream's properties and hence determine a realistic orbit.

\subsection{Small-angle approximation}
\label{subsec:small-angle}

\begin{figure}
\begin{center}
\includegraphics[width=0.35\textwidth, angle=0]{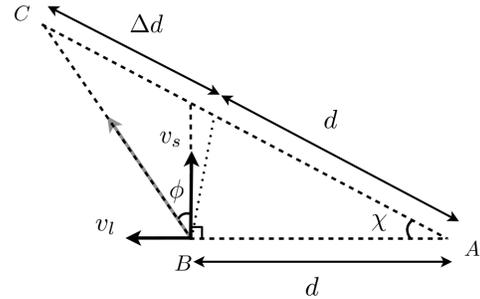}
\caption{Schematic diagram to illustrate how equation~(\ref{eqn:vs})
  can be reached in the small-angle regime for objects with a more
  limited angular extent than a generic stellar stream.  $A$ denotes
  the position of the Sun, $d$ is the heliocentric distance to the
  object at $B$, and $BC$ gives the true extent of the object with the
  object's elongation being along the grey arrow.  The difference in
  distance between the extreme ends of the object is given by $\Delta
  d$, where $\Delta d << d$.  Equation~(\ref{eqn:vs}) is recovered
  straight-forwardly when $\chi<<1<<\phi$ and the observed elongation
  of an object is assumed to be a simple projection of its true
  elongation onto the plane of the sky, with the orbit following the
  direction of elongation, such that $\tan\phi=v_l/v_s$
  .\label{fig:schematic}}
\end{center}
\end{figure}

The same relation for $v_s$ can be reached in a straight-forward
manner in the small-angle regime through simple geometric arguments,
when the object of interest is, for instance, a disrupting dwarf
galaxy becoming unbound and is thus no longer self-gravitating.  Such
a system would exhibit a large elongation and a radial velocity
gradient along its extent.  In these cases, one assumes that the
observed elongation can be de-projected to give the true orbital
direction, whose angular difference to the sky projection, $\phi$, is
given by the relation $\tan\phi = v_l/v_s$.  This implies that if the
transverse velocity can be estimated through other means, then the
variation in distance across the face of the galaxy can be easily
deduced \citep{Hercules}.  Figure~\ref{fig:schematic} illustrates how
equation~(\ref{eqn:vs}) may be reached in such cases, with the
pictorial representation showing quantities in the plane of the orbit.
Note that this figure is {\it{not}} to scale as, in reality, $\Delta
d/d << 1$.  In cases where $\chi$ is small (or, more specifically,
when $\chi<< 1 <<\phi$), equation~(\ref{eqn:vs}) is recovered by
applying the small-angle approximation directly to
Figure~\ref{fig:schematic} and assuming that the true elongation (in
the direction of the grey arrow) follows the velocity vector.  We also
outline below the steps taken to reach the same answer when leaving
the application of the approximation to the end; in this case, one can
initially employ the sine and cosine rules to obtain exact relations
between the various lengths and angles involved.

Let $L$ be the true extent of the object, denoted by $BC$ in
Figure~\ref{fig:schematic}.  Then:
\begin{equation}
\label{eqn:sine}
\frac{d}{\cos(\phi+\chi)} = \frac{L}{\sin\chi}\,,
\end{equation}
\begin{equation}
\label{eqn:cos_dd}
(d+\Delta d)^2 = d^2 + L^2 + 2dL\sin\phi
\end{equation}
and
\begin{equation}
\label{eqn:cos_d}
d^2 = (d+\Delta d)^2 + L^2 - 2(d+\Delta d)L\sin(\phi+\chi)\,.
\end{equation}
By taking the difference between equations (\ref{eqn:cos_dd}) and
(\ref{eqn:cos_d}), and then using (\ref{eqn:sine}) to eliminate $L$,
one arrives at equation~(\ref{eqn:vs}) in the limit that
$\chi<<1<<\phi$ and $\Delta d<<d$.  The latter assumption is only
necessary in so far as the fact that the assumed linear elongation,
$L$, should be a good representation of the direction of orbital
motion at that location.

\section{Summary}
\label{sec:summary}

This letter presents a simple result that enables the calculation of
the transverse velocity along a stream by using the gradient in
distance along it.  If at least one radial velocity measurement is
available and the distance gradient there is known, the transverse
velocity at this location can be calculated without the knowledge of
the Galactic potential. In these cases, one then has all of the
6-dimensional phase-space information necessary to derive the orbit
for the stream.  For cases where the radial velocity is not known, one
still gains knowledge of the ratio between the radial and transverse
velocity components.  It is therefore still possible to deduce
¯families of orbits that are functions of the initial radial velocity
alone.  This result is particularly useful for the study of Galactic
stellar streams, in cases where there is a lack of spectroscopic
information, but when ample photometric data are available for stellar
populations with reliable distance measurements and with small
associated errors.  With current and future large-sky photometric
surveys --- such as the SDSS and Pan-STARRS --- having the potential to
find a multitude of stellar streams in the Milky Way's halo, we hope
that the result presented here will also be useful as a first step in
the numerical studies of disrupting stellar substructures in the
Galaxy.

\subsection*{Acknowledgements}
We thank James Binney and Martin Smith for providing comments on a
draft version of this paper.  We also thank the referee, Paolo
Miocchi, for helpful comments.  SJ would also like to thank
Eric Bell for discussions which inspired this work.

\end{document}